\documentclass[authoryear,review]{article}

\makeatletter
\def\ps@pprintTitle{%
	\let\@oddhead\@empty
	\let\@evenhead\@empty
	\def\@oddfoot{}%
	\let\@evenfoot\@oddfoot}
\makeatother

\usepackage[utf8]{inputenc}
\usepackage{fontenc}[t1]
\usepackage{lscape}

\usepackage{geometry}                
\usepackage{amssymb}
\usepackage{amsmath}
\usepackage{amsthm}
\usepackage{bigints}
\usepackage{placeins} 
\usepackage{lineno}
\usepackage{adjustbox}

\usepackage{graphicx}
\usepackage{epstopdf}
\theoremstyle{plain}

\newcommand{\komment}[1]{}

\title{Contact Tracing -- Old Models and New Challenges }
\author{Johannes M\"uller$^{1,2}$, Mirjam Kretzschmar$^{3}$}

\date{
{\small $^{1}$ Mathematical Institute, Technische University M\" unchen, Boltzmannstr. 3,
	85748 Garching, Germany }\\
{\small $^{2}$ Institute for Computational Biology, Helmholtz Center Munich, 85764  Neuherberg, Germany}\\
{\small $^{3}$ Julius Center for Health Sciences and Primary Care, University Medical Center Utrecht, Utrecht University, Heidelberglaan 100, 3584CX Utrecht, The Netherlands}}

\begin{document}
	\maketitle
 \begin{abstract}
Contact tracing is an effective method to control emerging diseases. Since the 1980's, mathematical modelers are developing a consistent theory for contact tracing, with the aim to find effective and efficient implementations of contact tracing, and to assess the effects of contact tracing on the spread of an infectious disease. 
Despite the progress made in the area, there remain important open questions. In addition,  technological developments, especially in the field of molecular biology (genetic sequencing of pathogens) and modern communication (digital contact tracing), have posed new challenges for the modeling community. 
In the present paper, we discuss modeling approaches for contact tracing and identify some of the current challenges for the field.
 \end{abstract}
	
\section{Introduction}

Emerging and re-emerging infectious diseases like SARS, Ebola, Lassa fever, Tuberculosis, and most recently SARS-CoV-2,  require rapid responses and targeted control measures. It is best if an immunization of the population is possible - however, in case of emerging diseases, often the necessary vaccines are not available yet, or not available in sufficient quantities. An alternative approach is to stop infection chains by non-pharmaceutical control measures, such as reducing infectious contacts by social distancing, and testing and isolating infectious individuals. This can be done by screening if sufficient testing facilities are available, and in addition tracing and quarantining contact persons of infected index cases~\cite{klinkenberg2006}.\\
Mass screening as a stand-alone control measure is effective if prevalence is high, and cheap, rapid, and reliable diagnostic procedures are available. Otherwise, most tested persons are uninfected such that even a small probability for a false positive test result leads to a large number of false alarms, whereas only few infected persons can be identified.\\
Contact tracing (CT) is a more focused method: Once an infected individual is diagnosed and isolated, contact persons are identified who had potentially infectious interactions with that index case. In general, the prevalence within this group will be much higher than that in the general population, such that it may be effective to screen these persons, and to quarantine or isolate them. CT acts on several levels of the transmission process: 
{\it Individual level:}  Infected persons are diagnosed early and are isolated or receive medical help.  
{\it Population level: } Transmission chains can be detected and stopped, which reduces the effective reproduction number. 
{\it Medical/scientific level: } It is possible to study infector--infectee pairs, and to learn about who infected whom in the outbreak. This gives information on risk factors,  transmission modes, infectivity, and generation intervals of the infectious disease. This  information can be central in the design of further control measures.\par\medskip 
The main challenge in mathematical modeling of CT is the individual-based character of the process. Information about the health status of single individuals and the time course of the contacts  between those individuals is necessary. An appropriate formulation of the process is only possible at the microscopic level. However, the main interest is at the mesoscopic and macroscopic level: Not single individuals but the spread of the infection is in the focus of the interest. Model approaches need to span several scales, from the individual to the population. Similar to CT, also the transmission of infection is a process occurring between individuals. That process can be readily approximated by mean field models as the Kermack-McKendrick model. For the epidemic process, it is well established how to bridge the scales. CT, in contrast, follows the interaction given by transmission events, so can be viewed as a kind of superinfection: One could consider CT as an infection that follows the paths of the primary infection, and removes in that way infected individuals. This process superimposed on the transmission process creates a high degree of dependency between individuals. Therefore, lifting a model description for CT from the individual level to the population level involves more technical difficulties than for the case of a pure infection process.  
\par\medskip 

Intuitively, the CT and transmission processes are racing each other. Starting at an index case, the infection spreads to contacts, while with CT we aim to catch up. This picture already indicates many of the properties that make a disease controllable by CT~\cite{fraser2004}: (a) a sufficiently large fraction of cases develops symptoms and is tested and diagnosed; (b) contacts need to be well identifiable and tracable; (c) the disease spreads slowly enough to allow CT to catch up, even if identification, testing, and quarantining of contacts comes with a certain delay, and (d) diagnostic tests are able to readily identify symptomatic and asymptomatic persons.\\
Here, we review published literature on mathematical models for CT and their applications. We then identify and discuss open problems and current challenges for the theory of CT.

\section{Modelling approaches}
As a rigorous mathematical analysis of detailed models for CT is rather challenging, a multitude of simplifying approaches have been established. Among the first papers is a study by Hethcote and Yorke concerning interventions for gonorrhoea~\cite{hethcote:gonorrhea}. They include CT in their model and suggest that the effect of CT is a reduction of the transmissibility of the infection. In the same spirit,~\cite{Hyman2003} proposes that CT is based on the identification of infected persons by their infected contacts. In that, the ``superinfection'' is taken verbally, and infecteds are removed with a term that resembles the incidence term (product of discovered infecteds and prevalence). In those models, the term representing CT is not based on first principles, but rather defined {\it ad hoc}. \\

\noindent For models, that are based on first principles, three main directions can be identified: 
\begin{enumerate}
\item Simulation models that directly simulate individuals and large populations; 
\item Pair approximation models that are extensions of mean field models incorporating information about correlations of pairs of individuals;
\item Stochastic and deterministic models that are based on a rigorous analysis of a simple branching process modeling CT. 
\end{enumerate}

\noindent We discuss these in the following sections, and also the phenomenological approaches, which are relevant for practical applications. phenomenological approaches, which are relevant for practical applications.

\subsection{Individual based simulation models}
Individual (or agent) based models (IBMs)~\cite{Willem2017} are perhaps the first choice to formulate a process as complex as CT. Individual-based models describe the fate of every single individual in a population and their interactions. IBMs explicitly incorporate a contact graph, which can be as simple as a complete graph where every individual has contact with every other individual (fully connected population), a random graph as described by the configuration model, or a small world graph~\cite{Miller2011,keeling2005networks,Newman2001} that reflects local and long distance contacts. Conceptual, parsimonious  models are used to address the influence of the contact graph structure on transmission dynamics~\cite{kiss2006,kiss2008}.  The most detailed IBMs describe graphs that aim to represent  existing societies with cities, work places, and schools~\cite{Meyers2005}. It depends on the aim of the model how detailed the contact graph, the state of an individual, and its behavior is formulated. As such a model is algorithmic, there is almost no limit for the degree of detail that can be included. However, very detailed models are often faced with the problem of lack of data for an appropriate parametrization. \\
In any case, the advantages of individual-based models for CT are three-fold: First, they are simple to formulate and to communicate; second, they allow to directly represent CT in a realistic way and in that foster and guide the development of more abstract, analytical models; and third, they provide sufficient detail, such that results can be used in public health decision making.
\par\medskip 

IBMs are frequently used to investigate specific infectious diseases, and to extract relevant information about the effectiveness of intervention strategies. Contact tracing is commonly performed to improve case finding for sexually transmitted diseases as gonorrhoea and chlamydia; for these diseases, contacts can be clearly defined - recent sex partners - and a large fraction of infections are asymptomatic, such that contact tracing greatly enhances the possiblity of finding and treating infected persons. Also, the time scale of transmission of these infections is sufficiently slow to enable finding and treating contacts on a faster time scale than the generation interval. Finally, as sexual partnerships are often long lasting, treating contacts may prevent re-infections of the index case~\cite{Kretzschmar1996,ghani1997role,Turner2006,Kretzschmar2009, Kretzschmar2012, Althaus2012}. Several IBM models investigate  Tuberculosis~\cite{Tian2011,Tian2013,Kasaie2014,Mellor2011} (see also the review article~\cite{begun2013} and references therein), Smallpox~\cite{Porco2004}, also  in connection with bioterrorism~\cite{eichner2003}, measles~\cite{Liu2015}, and Ebola~\cite{Shahtori2018}. Also, the effect of CT on control of SARS has been simulated by various IBMs, see the e.g.~\cite{klinkenberg2006,Lloyd-Smith2003}, and particularly the review article~\cite{Kwok2019} and references therein. Peak et al.~\cite{Peak2017} compare effectiveness of CT for several infections, including Ebola, influenza, and SARS. In~\cite{Armbruster2007,Armbruster2007a}, CT is analyzed from a public health economics point of view. The effect of CT on the SARS-CoV-2 pandemic was investigated based on IBMs in~\cite{Bradshaw2020,Bulchandani2020,Hellewell2020,HernandezOrallo2020,Keeling2020,Kucharski2020,
Lorch2020,Tanaka2020,Peak2020}. 

\subsection{Pair approximation models}
Stochastic processes are hard to analyze, while many tools for deterministic models are available. Mean field equations are a well established, heuristic approach to reformulate individual-based models as described above in terms of ordinary differential equations (ODE's). In the limit, if the population size tends to infinity, the approximation becomes exact in case of a homogeneous population; that precisely is the way to derive 
a deterministic model as the Kermack-McKendrick (or SIR) model from a stochastic process for an infection. Instead of counting the number of individuals of different types, the expected relative frequencies are addressed by the ODE model. Some transitions, as recovery of an individual, are independent of the states of all other individuals. In that, the expectations exactly satisfy an ODE. 
Other transitions, as transmission of infection, are based on interaction of individuals. At that point, only an approximation is possible that neglects all correlations in the contact graph. Formally, the expectation of the product of random variables is replaced by the product of expectations~\cite[chapter 3.1.10]{mueller:kuttler}. Under appropriate conditions, particularly if the contact graph is a complete graph (contacts happen between any of the individuals), this procedure becomes rigorous if the population size becomes large or infinite. The success of ODE models to describe the time course of real-world epidemics is an {\it a posteriori} justification for simple deterministic models. In many cases, they allow for a deeper understanding of the underlying mechanism, and for powerful predictions~\cite{Keeling1999}.\par\medskip 
All information about correlations are lost in the transition from an IBM to a mean field model. In that, a mean field model is insufficient to appropriately describe the infectious process on an inhomogeneous contact graph, particularly if the contact graph is strongly locally clustered: Is this the case, the neighbor of an infected individual often already is infected, and the spread of infection slows down. In the 1990's, mainly driven by Japanese~\cite{Sato1994} and British~\cite{keeling1997correlation,Keeling1999} groups, an improved mean field approximation was developed, the pair approximation. Here, not only expectations but also the correlations are formulated in an ODE model. That is, not only the expected number of individuals in state $S$, $I$, or $R$, say, but also the expected number of edges connecting e.g.\ $I$ with $I$ or $S$ with $I$ individuals are followed in the system of ODE's. As the model incorporates information about correlation, it can -- up to a certain degree -- mimic the slow down of the spread caused by spatial correlations. The disadvantage of this approach is the exploding number of equations needed, as not only one equation per state is required, but also one for each pairwise combinations of two states. Therefore, most authors do not perform an analysis based on the theory of dynamical systems, but the resulting ODEs are solved numerically. 
Furthermore, it turns out that -- for strongly localizing graphs --  the conclusions are rather of qualitative than quantitative nature. If the contact graph is well mixing (high degree of the nodes, and no clustering), the pair approximation becomes better (which is also the case for the mean field approximation). Recent developments allow even an exact analysis by the ``message passing method'', a subtle further development of the pair approximation~\cite{karrer2010message,sharkey2013kj,wilkinson2014message} for trees.\\
Concerning the modeling of CT, pair approximation keeps a central piece of information that is dismissed by the mean field approximation: We know how likely it is that a neighbor of an infected individual is infected. In that, it is possible to remove infected neighbors of an index case~\cite{eames2002,Huerta2002}. This idea has been discussed in a series of papers~\cite{Tsimring2003,eames2003contact,keeling2005networks,house2010,house2011}. The comparison with Monte Carlo simulations of the stochastic process indicates that the results are valid especially for large, homogeneous graphs. Most of these papers investigate the effectiveness of CT in different contact graph structures. 
In ~\cite{house2010} it is found that CT is more effective in clustered  than in homogeneous populations. In~\cite{eames2007contact} recursive and one-step tracing is compared, and ``targeted CT'', that is, CT focusing on a risk group, is analyzed. Recursive and targeted CT were found to be particularly effective. 
The models for CT based on pair approximation are rather conceptual models that allow addressing fundamental questions than models used to quantitatively predict the effect of CT on the spread of  real-world infections. One of the few exception is~\cite{Clarke2012}, where pair approximation is applied to predict the impact of CT on chlamydia prevalence. A simpler approach, in which only pairs are taken into account, which may form and dissolve, is used in ~\cite{Heijne2011} to investigate the impact of screening and CT, again addressing the prevalence of chlamydia.

\subsection{Models based on branching processes}
At the onset of an outbreak spreading in a homogeneous population, it is possible to replace the epidemic process by a birth-death process of independent individuals, where ``birth'' means a new infection and  ``death'' recovery. Particularly the interaction of two infectious individuals is unlikely and negligible in a large, homogeneous population~\cite{ball:donnelly}. The statistics of this process, e.g.\ the size and structure of connected components is known~\cite{mueller2003}. On top of this linear birth-death process, a process of CT can be formulated. It is possible to rigorously analyze this stochastic process~\cite{mueller2000, Ball2011}: CT mainly affects the removal rate of infected individuals. That is, in order to address the stochastic process, the probability to be infectious at a given time after infection is determined. This probability is the central function that allows to readily determine the effective reproduction number, or the doubling time of an infection. Various aspects of CT can be investigated in this context, as the recovery and infectivity depending on the time since infection~\cite{mueller2000,Ball2011}, estimation of the tracing probability from data~\cite{mueller2007,Blum2010}, or the effect of a tracing delay~\cite{Ball2015,muller2016effect}. Strictly spoken, the analysis and the results are valid only for the onset (or during the decay before exinction) of the outbreak, if direct contacts between infecteds are unlikely to happen. However, using heuristic arguments, the removal rate can be approximated also in the case of high prevalence, and a modified mean field equation has been proposed~\cite{mueller2000}.

 Similar to pair approximation, the approach is not suited for a complex contact graph structure with small homogeneous clusters that only weakly interact. It is interesting that the central idea for the analysis of CT on the one hand, and the message passing methods used in recent versions of pair approximation on the other hand, bear a remarkable similarity. In~\cite{Okolie2020}, the branching process analysis is generalized from homogeneous populations to populations with a prescribed contact graph, and ideas are discussed how to merge the branching process analysis and the pair approximation. Also~\cite{Kojaku2020} investigates CT on a random graph, this time via an approximation using generating functions for he degree distribution of neighbors of randomly chose infected individuals. They find, that the degree of an individual detected by backward tracing roughly behaves as the expectation of the squared degree, indicating the high efficiency for CT to detect the persons that are best connected.\\
Brown et al.~\cite{Browne2015} developed a sophisticated ODE-approximation of the branching-process structure and applied that to Ebola. Becker et al.~\cite{Becker2005} also proposed a simplified version of the model and investigated the SARS epidemic in a deterministic model with household-structure. Kretzschmar et al.~\cite{Kretzschmar2004} used a branching process to model ring vaccination for smallpox. This model was recently developed further to investigate the effectiveness of CT for SARS-CoV-2~\cite{Kretzschmar2020,Kretzschmar2020a}.
Based on a branching-process formulation for CT, Tanaka~\cite{Tanaka2020} analysed data for SARS-CoV-2 to estimate the fraction of asymptomatic cases.

\subsection{Phenomenological approaches}
Phenomenological approaches are not rooted in an analysis of stochastic processes that obviously model CT in an adequate way. In that, it is difficult to understand if the terms chosen to model CT are adequate. The advantage of these models is their simplicity -- they are ODE models with rather simple structure (concerning CT) and can be readily analyzed or simulated. In that, these models are suited  to address real world epidemics. The difficult part is the interpretation of the results (again concerning CT), as the clear connection with first principles is not obvious. Often, the tracing part is either formulated as a linear removal term, or as a mass action term.
\par\medskip 
Many of these models are applied to the HIV infection, where mostly a mass action term is used to represent CT
\cite{Arazoza2002,Hyman2003,Hyman2003,Clemencon2008,Naresh2011,AGARWAL2012}, but sometimes also simply a linear term~\cite{Hsieh2010}. In ~\cite{Clemencon2008,Hsieh2005} several ways to model CT -- linear, mass action, and a saturation function -- are compared with data. Interestingly enough, results in~\cite{Hsieh2005} indicate that a mass action term for CT is inferior to a linear term or a saturation function. In a similar spirit, Clarke et al.~\cite{Clarke2013} adjust a power low term for CT based on simulations from an IBM. 
Also models for Chlamydia~\cite{Heffernan2009}, Tuberculosis~\cite{Aparicio2006}, Smallpox~\cite{Kaplan2002}, Ebola~\cite{Berge2018}, and models for SARS-CoV-2~\cite{Giordano2020,Lunz2020} are based on this phenomenological modeling approach. 
\par\medskip 
Fraser~\cite{fraser2004} proposed an idea with a slightly more profound connection to CT. The model is based on age-since-infection. In that, it is similar to the branching-process analysis, but CT is formulated as a linear effect. In that article, characteristics in the timing between onset of symptoms and infectivity are identified that make an infection controllable by CT. Chen et al.~\cite{Chen2006} took up that approach to analyze a model for the SARS infection and Ferretti et al.~\cite{Ferretti2020} applied the model to the SARS-CoV-2 epidemic in Italy.

\section{Challenges}
The efforts of the last  40 years to develop a toolbox for CT  have delivered a considerable amount of modelling approaches and results. There is a general agreement about a fundamental model structure, that can be readily realized in IBMs, and there are various ways to analyze the stochastic process either approximately or rigorously -- at least, if we stick to simple models. Many simple {\it ad-hoc} models are published. Nevertheless, some of the central questions are still open.

\subsection{Classical questions}
Some questions and problems have been debated for quite a while. We pick a few of these ``classical questions'' which we consider as interesting and/or of practical need, and discuss them.\par\medskip 

{\it Modeling CT.} Obviously, a multitude of models are in use. Some models, as the stochastic IBMs, directly simulate CT. In an IBM, it is easy to incorporate CT appropriately. Stochastic simulation models have the advantage that their outcomes are easy to interpret, but the disadvantage that they cannot be analyzed analytically. To obtain insight into the parameter dependence of CT and to develop general rules for CT is not straightforward or may be impossible  if we exclusively rely on IBMs.\\ 
Other approaches, as the pair approximation models or models based on branching processes, use first principles to develop the model structure. Therefore, also these models are able to reflect   CT in an appropriate manner. Even if approximations are used to derive simplified mathematical structures, it is straightforward to check their accuracy (compare the analytical results to simulations of the original models). In that, these models are appealing. However, in some cases, the derivation  is rather technical and not straightforward to communicate. This class of models are well suited for theoretical considerations. For practical purposes, simpler model structures seem to be desirable.\\
At that point, the phenomenological approach comes in. Here, the models are mostly compartmental (ODE) models, as often used in ecology and epidemiology. These models can be readily analyzed and simulated. In that, practical applications of such simple models are easily possible, which is the strength of the phenomenological approach. The drawback is the fact that they are not rooted in first principles. It is hard to assess whether the model structure appropriately reflects reality. E.g., if CT is formulated as a linear term, this formulation is contradicting the observation that CT is based on correlations and dependencies between index cases and their infectees/infector (as a rule, dependencies are expressed by nonlinear terms). Another aspect is the parametrization of the models: Some central parameters can be obtained by observations on the micro scale. E.g., the probability to detect a contact can be estimated using the number of detected cases per index case~\cite{mueller2007,Blum2010}. As phenomenological approaches dismiss the micro-scale and directly jump to the macro-scale, information that is available on the micro-scale is almost impossible to incorporate.\\
Easy to use models that are well accepted and approved by the modeling community are necessary but not yet in sight.
\par\medskip

{\it Backward/Forward tracing.} Even in one of the very first papers about CT, the seminal work by Hethcote and Yorke~\cite{hethcote:gonorrhea}, the distinction between backward and forward tracing is mentioned. Backward tracing means that the infector is detected by an infectee who becomes an index case, while in forward tracing the infector is the index case, while the infectee is detected. Since this first article, the relative importance of backward- and forward tracing is under discussion. An individual only has one infector, but in general several infectees. This fact might indicate that forward tracing is more important. On the other hand, if we randomly select an individual in a natural contact graph, the neighbor of this individual will on average have more contacts than an average individual. This finding, also called the ``friendship paradox'' in the context of social networks, has been proposed to be used in an early warning system for influenza~\cite{Christakis2010}: Students were asked to name two friends. The authors of the study monitored the occurrence of influenza among those named persons. It turned out that the incidence of influenze increased two weeks earlier than it did among the average students. This observation indicates that backward tracing also is of importance, as most likely we will find persons who have many contacts and already had the chance to infect many of them. This is especially important for CT in sexually transmitted infections, which are often circulating in highly connected core groups.\\ 
The question of whether there are super-spreaders also falls within this context~\cite{Lloyd-Smith2005,Galvani2005,Melsew2019}. The frequency and the importance of super-spreaders for the dynamics of infection, and also for the impact of CT, has not yet been conclusively clarified;~\cite{klinkenberg2006,Kojaku2020} argue verbally that backward tracing is readily able to identify super-spreaders, s.t.\ a combination of backward- and forward tracing is efficient, also in the presence of super-spreader events. In~\cite{Okolie2020}, quantitative comparisons (based on analytic results for a branching-process model for CT on random tees)  indicates that the effect of CT decreases with the variance of the degree distribution. This, in turn, is a hint that CT performs better without super-spreader events. Also simulation studies~\cite{Kiss2007} point in that direction (at least if we consider the expected final size). The timing (latency and incubation period) of the infection itself will also have some influence. There is a need to investigate the effect of super-spreaders on CT more in depth.   \par\medskip

{\it Endemic equilibrium.} CT is also performed if the infectious process is in its endemic equilibrium. In this case, in average each infected individual  only has one infector and one infectee ($R_{eff}=1$). Why does CT pay in that situation? For sexually transmitted diseases (STD), it is clear that the partner is of high risk, and a couple should rather be considered as a single  entity. Partner notification in faithful pairs is a simple case of CT. It is more interesting to note that for some STDs asymptomatic persons can be infectious for a long time (several months).  In that, particularly the tracing of asymptomatic persons may be decisive. CT is a method to find these highly infectious persons with possibly many contacts, who are difficult to localize otherwise. \par\medskip 

{\it Tracing probability.} In order to monitor the effect of CT, it is desirable to estimate the tracing probability, that is, the fraction of identifyable contacts among all infectious contacts. While data for the number of detected cases are available, the number of missed cases is usually unknown. In ~\cite{mueller2007,Blum2010}, some statistical methods are developed to estimate the tracing probability. A related problem is the estimation of the abundance of asymptomatic cases from tracing data. 
\cite{Dyson2017} proposes a method based on household models for 
Yaws, a disabling bacterial infection, while Tanaka~\cite{Tanaka2020} aims to estimate the percentage of asymptomatic cases for SARS-CoV-2 using the branching process approach for CT. However, these questions are rather neglected by the recent literature and deserve a deeper investigation.\\

\subsection{CT and genetic sequence data}
In recent years, genetic sequencing of pathogen DNA became rather cheap, and genetic sequence data are readily available. Methods of population genetics are used to, e.g., estimate the prevalence of an infection~\cite{Frost2013}. Genetic sequence data is also used to identify clusters of infections, and even to identify and refine transmission trees (infector/infectee relations) within a cluster~\cite{Roetzer2013,Pasquale2018,Dennis2018,Pasquale2020}. However, the combination of data from CT and from sequencing techniques is hardly exploited by now.  It is natural to ask what epidemiology could gain from that combination. \\
Clearly, from samples of infector-infectee pairs, the mutation rate of the pathogen can be estimated. Moreover, as an infection event forms a bottle neck for the population of the pathogen, the time since infection can be estimated, and in that, it is possible to narrow down the time of the infectious contact. A comparison of different infector/infectee pairs might help to sharpen the estimations for the prevalence. \\
The usage of genetic sequence data together with methods from population genetics is rather young, and many powerful methods -- as the  SMC~\cite{McVean2005} -- are rather recent developments. We expect that useful tools become available in the near future. 

\subsection{Digital CT}
The idea to use data from mobile phones to trace contacts is rather recent. First practical attempts to use digital sensors (RFID chips) to observe contacts and to investigate an empirically validated contact network in relatively small communities (school, hospital, conference) reach back to 2010 \cite{Salathe2010,Isella2011,Stehle2011}. Soon it became clear that risk evaluation based on mobile phone data is possible and useful~\cite{Lima2015,Mastrandrea2016}. Perhaps the first paper that considered digital CT (DCT) was an IBM-based simulation study by Farrahi et al.\ in 2014~\cite{Farrahi2014}.  Several manuscripts in the context of SARS-CoV-2 also address  DCT~\cite{Bradshaw2020,Bulchandani2020,Lunz2020,Ferretti2020,HernandezOrallo2020,Kretzschmar2020a}.\\
The ability to localize persons using data generated by smartphones already detects traffic jams~\cite{Pan2013}, and is a source of major concern as these data could also be misused~\cite{Taylor2015}. DCT could help to improve some crucial shortcomings of classical CT. The advantages of DCT are the rapid identification of infectious contacts, and the possibility to also rapidly inform contactees. In comparison with conventional CT, DCT has the potential to strongly reduce the tracing delay, and to increase the tracing probability. In that, infections may become controllable that were not controllable before. However, these potential benefits come with a number of technical, social, medical, and practical challenges. Contacts identified by mobile devices need to correlate with infectious contacts. A rapid and cheap test for the infection is necessary, as the number of persons to test will be much higher than in conventional CT. Aspects of privacy and data protection are crucial elements of DCT. Apps need to be accepted in the society. The concept of DCT, its strengths and its weaknesses have to be communicated clearly, in a way that citizens understand and accept DCT. The system can only work if a large part of the population participates in DCT. \\
In the present note, we rather focus on new modeling challenges posed by DCT, and less on the social and technical aspects, though all of these points are intertwined.\par\medskip 

{\it Correlation between individuals.} Often, the mathematical analysis of CT models focus on the onset of the outbreak. In that, correlations (particularly between infected individuals) that do not come from an infector/infectee relation can be neglected. At present, conventional CT can afford to trace contacts only if rather few index cases are diagnosed. Once the number of diagnosed infected persons exceeds several hundred in a community, the effort to identify manually many contacts per person is not feasible anymore.\\
DCTS promises to overcome some of the logistic problems of CT and to allow for the identification of a huge number of contacts, even if a high number of infected individuals are present. The number of reported contacts per day and person in Europe is in the magnitude of 10~\cite{Mossong2008}, with a large standard deviation (which is in the same range as the average number of contacts). If we have a tracing window of one week, we easily estimate 70 contact persons per index case (with a high variance). Depending on the nature of the infection, we can (and should!) immediately inform not only the direct contactees but also the contactees of the contactees (second level tracing). Note that second level tracing is dissimilar to the classical two-step tracing: In two-step tracing a contactee is first tested, and if the test is positive, he/she becomes an index case and the snowballing continues, while in second level tracing the contactees of contactees are immediately informed about their possible exposure, without waiting for a test result of contact persons. In that, we easily arrive at 500 direct and indirect contactees per index case. That is, the fraction of the population affected by DCT is around 500 times larger than the number of index cases, however, the number of actually exposed persons might be small. From a practical aspect, it is of utmost importance to identify features of an infection that imply the necessity of first- second- or even higher-level tracing.\\ 
The difference to the conventional procedure is the possibility for rapid and immediate information of direct and indirect contactees, without waiting for medical tests and diagnoses.  Even in a moderate outbreak, it is likely that the groups of contactees related to different index cases will overlap. These effects tremendously complicate the mathematical analysis of the models. A clear challenge is the identification of techniques that allow for the mathematical analysis of this situation.   
\par\medskip 

{\it High number of contactees / practical protocol.} As discussed above, DCT can identify a high number of contractees and inform them about direct or indirect infectious contacts. From practical considerations, a reduction of that number is desirable. Technical devices may give a score to each contact, and estimate the probability for infection. That might help, but it also might be the case that scores are not reliable. \\
Most likely, we are faced with a serious practical problem: If all casual contacts are reported, many persons might receive  a warning rather often. In that case, fatigue sets in and warnings are not taken seriously anymore. The DCT becomes useless. A possibility to escape this risk is the choice of a smaller subgroup from the set of contactees. One possibility could be simply a random sub-set. More effective is a classification of contactees in different risk groups, e.g. according to the number of recent contacts. Particularly asymptomatic persons that spread the infection for a longer time will appear among repeatedly identified contactees.\\
Modeling approaches need to clarify the situation and to estimate the number of warnings a person receives under realistic conditions. In consequence, different strategies of reducing the number of contactees to be informed have to be defined. Models allow to estimate the impact of these strategies, and to filter out reasonable policies for DCT. Many parameters will influence the outcome, as the availability of tests, the social/contact structure, and the abundance of a tracing app on mobile phones. \par\medskip

{\it DCTS induce an inhomogeneity in the population.} The population is divided in a subpopulation with, and a subpopulation without a tracking device. In that, contacts within the DCTS-subpopulation are readily identified, while contacts between the two subpopulations or among the non-DCTS-subpopulation can only be identified by conventional CT. That situation is completely different from an imperfect test, which leads to a certain probability of undetected infectious contacts. It is likely that the acceptance of a tracing device correlates with social factors as education, political, or religious orientation. In that, connected subgroups appear that escape the detection by DCTS. These subgroups might form a reservoir for the infection, from which importation to the general population may occur. The fraction of the population equipped with DCTS-equipment alone is not decisive, but also the distribution in a heterogeneous population. However, even for a homogeneous population, the division into two subgroups might lead to unforeseen effects. This aspect is a new one that deserves deeper investigation.

\section{Conclusion}
The joint effort of modeling and theoretical investigation of CT has borne many fruits, as the estimation of the impact of CT on the spread of infections, or methods to assess the influence of the tracing delay.  Nevertheless,  questions of practical importance still deserve our attention. New challenges, e.g.\ due to developments in genetic sequencing of pathogens and DCTS, have  arisen. Simulation models promise first answers on a rapid time scale, but it will take some time until these new aspects are fully understood analytically, and this understanding is translated into practice.

\addcontentsline{toc}{section}{References}
\bibliographystyle{abbrv}
\bibliography{ctBib}

\newpage 
\begin{appendix}

\section{Literature overview}
The table orders the papers primarily according to the infection they aim to investigate resp.\ for the rather theoretical papers, by the method used. \\

\noindent
\begin{adjustbox}{width=\textwidth,center}		
	\begin{tabular}
		{p{2cm}|p{1.5cm}|p{2cm}p{2cm}|p{0.5cm}|p{3.0cm}|p{4.0cm}}
		author&infection&method&term for CT&CT delay&remarks&outcome(w.r.t. CT)\\
		\hline
		Becker et al.\ 2005~\cite{Becker2005}&SARS&phenomen. / next generation operator&reduction of incidence& - &article exclusively aims at Reff, no dynamics.&formula for Reff; social distancing together with CT can control SARS\\
		Chen et al.\ 2006~\cite{Chen2006}&Influenza, Measles, Chickenpox,SARS&phenomen. / PDE& fixed fraction of newly infecteds are eventually traced & - &based on Fraser~\cite{fraser2004}; only Reff, no dynamics; airborne infection&Probability to control an outbreak is estimated\\
		Kwok et al.\ 2019~\cite{Kwok2019}&SARS&&&&review article&\\
		Fraser et al.\ 2004~\cite{fraser2004}&Theory \& Influenza, SARS, Smallpox,HIV&phenomen. / PDE& fixed fraction of newly infecteds are eventually traced& - &basic model, used in applications~\cite{Chen2006,Ferretti2020} &timing of incubation period and latency period is central for CT\\
		Lloyd-Smith et al.\ 2003~\cite{Lloyd-Smith2003}&SARS&phenomen. / time-discrete stoch. sim on pop.-level&increased transition rate to quarantine & + &CT not explicitly formulated&crucial that CT is implemented at the beginning of the outbreak\\
		\hline
		Bradshaw et al.\ 2020~\cite{Bradshaw2020}&SARS-CoV-2&IBM, homogeneous population&direct formulation& + &preprint; DCT considered, focus on onset, Reff and prob.\ for major outbreak&backward tracing and high abundance of DCT devices necessary\\
		Bulchandani et al.\ 2020~\cite{Bulchandani2020}&SARS-CoV-2&IBM, homogeneous population&direct formulation& - &preprint; DCT considered, focus on onset, heuristic formula for Reff&High DCT-device coverage necessary\\
		Ferretti et al.\ 2020~\cite{Ferretti2020}&SARS-CoV-2&phenomen. / PDE& fixed fraction of newly infecteds are eventually traced& + &based on Fraser~\cite{fraser2004}; DCT considered, heuristic formula for Reff&SARS-CoV-2 controllable by DCT\\
		Giordano et al.\ 2020~\cite{Giordano2020}&SARS-CoV-2&ODE/SIR&increased rate for removal& - &detailed ODE model &CT is a central element in controlling the infection\\
		Hellewell et al.\ 2020~\cite{Hellewell2020}&SARS-CoV-2&IBM, homogeneous population&direct formulation& - &onset of epidemic considered&high tracing probability necessary to control the infection\\
		Hernandez-Orallo et al.\ 2020~\cite{HernandezOrallo2020}&SARS-CoV-2&IBM, inhomogeneous population + ODE&direct formulation (IBM) / fixed fraction of newly infecteds are eventually traced (ODE) & + &simulations based on empirical contact network & CT needs to be precise in order ot avoid many persons in quarantine \\
	\end{tabular}
\end{adjustbox}

\begin{table}
	\begin{adjustbox}{width=\textwidth,center}		
		\begin{tabular}
			{p{2cm}|p{1.5cm}|p{2cm}p{2cm}|p{0.5cm}|p{3.0cm}|p{4.0cm}}
			author&infection&method&term for CT&CT delay&remarks&outcome(w.r.t. CT)\\
			\hline
			Keeling et al.\ 2020~\cite{Keeling2020}&SARS-CoV-2&IBM, homogeneous population&direct formulation& - &preprint; time and intensity of contacts vary, simulation-based estimation of Reff  &high tracing probability necessary to control the infection\\
			Kim et al.\ 2020~\cite{Kim2020}&SARS-CoV-2&Stochastic branching process&verbally& - &preprint; DCT considered; Heuristic calculations of efficiency&heuristic formula for efficiency\\
			Kretzschmar et al.~\cite{Kretzschmar2020} &SARS-CoV-2&IBM, homogeneous population&direct formulation& + &preprint; Model based on Kretzschmar(2004)~\cite{Kretzschmar2004}&middle range tracing probability necessary to control diseases\\
			Kretzschmar et al.~\cite{Kretzschmar2020a}&SARS-CoV-2&IBM, homogeneous population&direct formulation& + &preprint;  Model based on Kretzschmar~\cite{Kretzschmar2004}; DCT considered, combined with conventional CT& tracing delays need to be minimized for effective CT; DCT might be a way to speed up the process\\
			Kucharski et al.~\cite{Kucharski2020}&SARS-CoV-2&IBM, inhomogeneous population&direct formulation& - &preprint; DCT considered, combined with conventional CT&CT more efficient than mass testing\\
			Lorch et al.~\cite{Lorch2020}&SARS-CoV-2&IBM with discrete spatial structure& one-step tracing, if at similar times in the same location& - & DCT considered, combined with conventional CT&DCT efficient particularly in case of a low fraction of quarantined persons\\
			Lunz et al.~\cite{Lunz2020}&SARS-CoV-2&phenomen. / ODE SEIR&removal rate, mass action& - &preprint; the tracing rate is connected to contact heterogeneity but not based on first principles&optimal CT defined as minimizing the total number of individuals that go into quarantine during he outbreak\\
			Tanaka et al.~\cite{Tanaka2020}&SARS-CoV-2&IBM, homogeneous population&direct formulation& - &simulation of clusters detected by CT as input for stats (see also Blum~\cite{Blum2010})&Bayesian parameter estimation based on CT\\
			\hline 
			Berge et al.\ 2018~\cite{Berge2018}&Ebola&phenomen. / ODE, SEIR&  fixed fraction of newly infecteds go to quarantine& - &CT not explicitly formulated&stationary states and their stability analyzed\\
			Browne et al.\ 2015~\cite{Browne2015}&Ebola&branching process/ODE, SEIR& fixed fraction of newly infecteds go to quarantine& + &fraction of detected cases is computed based on the branching-process&paper aims at a theoretical framework that is feasible for practical applications\\
			Shahtori et al.\ 2018~\cite{Shahtori2018}&Ebola&IBM, homogeneous population&direct formulation& + &onset of infection only&crucial that CT is implemented at the beginning of the outbreak\\
			Rivers et al.\ 2014~\cite{rivers2014}&Ebola&phenomen., stoch. sim on pop.-level.+ODE, SEIR&increased diagnosis rate& - &CT not explicitly formulated&reduction of Reff by around 30\% possible\\
			\hline 
		\end{tabular}
	\end{adjustbox}
\end{table}

\begin{table}
	\begin{adjustbox}{width=\textwidth,center}		
		\begin{tabular}
			{p{2cm}|p{2cm}|p{2cm}p{2cm}|p{0.5cm}|p{3.0cm}|p{4.0cm}}
			author&infection&method&term for CT&CT delay&remarks&outcome(w.r.t. CT)\\
			\hline
			de Arazoza et al.\ 2002~\cite{Arazoza2002}&HIV&phenomen. / ODE SI&mass action& - &three classes of infecteds modeled: infecteds do /do not know their infection, AIDS&stability analysis of stationary points, comparison with data\\
			Cl{\'{e}}men{\c{c}}on et al.\ 2008~\cite{Clemencon2008} &HIV&phenomen. / stoch. process (population level), SDE, ODE & several terms: linear, mass action, saturation function & - &extension of de Azoza~\cite{Arazoza2002}&development of statistical tools (maximum likelihood estimators)\\
			Blum et al.\ 2010~\cite{Blum2010}&HIV&phenomen. / stochastic model on population level&linear and saturation& + & see also~\cite{mueller2007,Tanaka2020}  &Bayes interference (ABC and Metropolis Hastings) for estimating ct probability\\
			Hsieh et al.\ 2005~\cite{Hsieh2005}&HIV&phenomen. / ODE SI&several terms: linear, mass action, saturation function& - &extension of de Azoza~\cite{Arazoza2002}&mass action inappropriate, linear or saturation term for CT better\\
			Hsieh et al.\ 2010~\cite{Hsieh2010}&HIV&phenomen. / ODE SI&saturation function; two-step tracing& - &extension of de Azoza~\cite{Arazoza2002}&stability analysis of stationary points, Reff; two-step tracing superior over one-step tracing\\
			Hyman et al.\ 2003~\cite{Hyman2003}&HIV&phenomen. / ODE SI&mass action& - &two models: core group, different stages of HIV&Reff, sensitivity analysis\\
			Mellor et al.\ 2001~\cite{Mellor2011}&HIV and Tuberculosis&IBM with household structure&screening the household& - &casual contacts are not traced; HIV and Tuberculosis at the same time&cross-tracing of HIV and Tuberculosis is effective\\
			Naresh et al.\ 2006~\cite{Naresh2011}&HIV&phenomen. / ODE SI&  fixed fraction of newly infecteds know their infection& - &CT is not triggered by diagnosis, but infections of known infecteds&stability analysis of stationary points\\
			\hline 
			Clark et al.\ 2012~\cite{Clarke2012}& Chlamydia & pair approx. & removal rate on infected-diagnosed pairs & - & based on~\cite{eames2002, house2010} & CT is efficient particularly below a certain prevalence\\
			Clark et al.\ 2013~\cite{Clarke2013}& Chlamydia & IBM + phenomen. ODE SI & IBM: direct; ODE: power law in $I$& -&power law adapted to IBM simulations; optimal resource allocations (CT/screening) & CT is efficient and uses resources efficient\\
			Heffernan et al.\ 2009~\cite{Heffernan2009}&Chlamydia & phenomen. / ODE SI & mass action law & - & model includes random screening (yield index cases) and CT& model results in line with data\\
			Hethcote et al\ 1984~\cite{hethcote:gonorrhea}& Gonorrhea& phenomen.\ ODE & decreased incidence & - & introduce backward and forward tracing & pioneering work about CT\\
			Kretzschmar et al.\ 1996~\cite{Kretzchmar1996}&Gonorrhea, Chlamydia&IBM with household structure& identification of a fraction p of partners& - &one-step tracing&prevalence for different control scenarios\\
			hline 
		\end{tabular}
	\end{adjustbox}
\end{table}

\begin{table}
	\begin{adjustbox}{width=\textwidth,center}		
		\begin{tabular}
			{p{2cm}|p{2cm}|p{2cm}p{2cm}|p{0.5cm}|p{3.0cm}|p{4.0cm}}
			author&infection&method&term for CT&CT delay&remarks&outcome(w.r.t. CT)\\
			\hline
			Kretzschmar et al.\  2009~\cite{Kretzschmar2009} & Chlamydia & IBM & direct formulation & & three different IBM models previously published by different authors are compared & the results of the models are somewhat different, due to their complexity \\
			Turner et al.\ 2006~\cite{Turner2006}&Chlamydia&IBM&direct formulation &+& model with pair formation, CT within pairs & Model fits data, and yields comparable results comparable studies\\
			\hline
			Aparicio et al.\ 2006~\cite{Aparicio2006}&Tuberculosis&phenomen. / ODE SEIR& fixed fraction of newly infecteds are identified& - &CT not explicitly formulated&simulation of prevalence \\
			Begun et al.\ 2013~\cite{begun2013}&Tuberculosis&&&&review article&\\
			Kasaie et al.\ 2014~\cite{Kasaie2014}&Tuberculosis&IBM with household structure&screening the household& - &contacts outside the household are not traced&household tracing reduces the incidence by 2\%-3\%\\
			Tian et al.\ 2011~\cite{Tian2011}&Tuberculosis&IBM, inhomogeneous population&direct formulation& + &different scenarios, sensitivity analysis&simulation of prevalence \\
			Tian et al.\ 2013~\cite{Tian2013}&Tuberculosis&IBM, inhomogeneous population&direct formulation& + &Follow-up of Tien~\cite{Tian2011}&simulation of prevalence \\
			\hline 
			Agarwal et al.\ 2012~\cite{AGARWAL2012}&Influenza&phenomen. / ODE SIR&fraction of newly infecteds go to quarantine& - &two risk classes in susceptibles are considered&Dynamical systems analysis, Reff\\
			\hline 
			Eichner 2003~\cite{eichner2003}&Smallpox&phenomen. / Stochastic model on population level& all close contacts and fraction of casual contacts are traced& - &age of infection included in the model&critical tracing probability estimated\\
			Kaplan et al.\ 2002~\cite{Kaplan2002}&Smallpox&phenomen. / ODE& saturation function& - &contactees who are traced are vaccinated&mass vaccination superior to vaccination triggered by CT\\
			Kretzschmar et al.~\cite{Kretzschmar2004} &Smallpox&Stochastic branching process&direct formulation& + &ring vaccination triggered by contact tracing&delay in CT is crucial\\
			Porco et al.\ 2004~\cite{Porco2004}&Smallpox&IBM with household structure&direct formulation& + &one step and two-step tracing compared&massive CT and ring vaccination can control the outbreak.\\
			\hline 
			Liu et al.\ 2015~\cite{Liu2015}&Measles&IBM, inhomogeneous population&direct formulation& + &complex/realistic contact structure&CT can significantly contribute to control a measles outbreak\\
			\hline 
			Ball et al.\ 2011~\cite{Ball2011}&Theory&Stochastic branching process&direct formulation& - &SIR, focus on fixed and exponentially distributed infectious period&analytic approach, bounds on Reff, extinction probability\\
		\end{tabular}
	\end{adjustbox}
\end{table}

\begin{table}
	\begin{adjustbox}{width=\textwidth,center}		
		\begin{tabular}
			{p{2.2cm}|p{1.5cm}|p{2cm}p{2cm}|p{0.5cm}|p{3.0cm}|p{4.0cm}}
			author&infection&method&term for CT&CT delay&remarks&outcome(w.r.t. CT)\\
			\hline
			Ball et al.\ 2015~\cite{Ball2015}&Theory&Stochastic branching process&direct formulation& + &SEIR, follow up of Ball~\cite{Ball2011}&effect of tracing delay, Reff, extinction prob.\\
			Müller et al.\ 2000~\cite{mueller2000}&Theory&Stochastic branching process&direct formulation& - &focus on age since infection &Reff, ODE approximation, critical tracing probability\\
			Müller et al.\ 2007~\cite{mueller2007}&Theory&Stochastic branching process&direct formulation& - &based on Müller~\cite{mueller2000},  see also~\cite{Blum2010,Tanaka2020}&estimation of tracing probability\\
			Müller et al.\ 2016~\cite{muller2016effect}, &Theory&Stochastic branching process&direct formulation& + &based on Müller~\cite{mueller2000}&effect of tracing delay and latency period, Reff\\
			Okolie et al.\ 2018~\cite{Okolie2020}&Theory&Stochastic branching process&direct formulation& - &connects branching process and pair approx., based on Müller~\cite{mueller2000}&effect of a random contact graph on CT\\
			Klinkenberg et al.\ 2006~\cite{klinkenberg2006}&Theory \& influenza, smallpox, SARS, and foot-and-mouth disease&Stochastic branching process&direct formulation& + &single-step and recursive tracing&mostly: single step and recousive tracing is equal effective\\
			Shaban et al.\ 2008~\cite{shaban2008}&Theory&Stochastic branching process&direct formulation& - &Vaccination of detected individuals&Reff, probability for extinction, simulation of final size\\
			Kojaku et al. 2020~\cite{Kojaku2020}&Theory&Stochastic branching process&direct formulation&-& CT on random graph, focus on generating functions for the degree& CT highly efficient as nodes with high degree are traced\\
			\hline 
			Kiss et al.\ 2006~\cite{kiss2006}&Theory&IBM, inhomogeneous population&direct formulation& - &isolation of susceptible; scale free and Poisson network&For scale free networks, tracing effect less sensitive to the epidemiological parameters\\
			Kiss et al.\ 2008~\cite{kiss2008}&Theory&IBM, inhomogeneous population&direct formulation& - &assortatively / disassortatively mixing networks; Single-step and recursive tracing &CT more effective in dissacociative networks; recursive tracing more efficient\\
		Farrahi et al.~\cite{Farrahi2014}& Theory & IBM, inhomogeneous population&direct formulation& + & first model for digital CT & digital CT can be efficient even if the fraction of app-users is small\\
			\hline
			Eames et al.\ 2002~\cite{eames2002}&Theory / STD&pair approx.\& stoch sim.&direct formulation& - &this paper introduced pair approximation for CT&modeling CT by pair approximation\\
			Eames et al.\ 2003~\cite{eames2003contact}&Theory&pair approx.\& stoch sim.&direct formulation& - &based on Eames~\cite{eames2002}&critial tracing probability\\
			Eames et al.\ 2005~\cite{keeling2005networks}&Theory&pair approx.\& stoch sim.&direct formulation& - &based on~\cite{eames2002}; focus on different social graphs (small world, scale-free)&network structure influence efficiency of CT\\
		\end{tabular}
	\end{adjustbox}
\end{table}

\begin{table}
\begin{adjustbox}{width=\textwidth,center}		
\begin{tabular}
				{p{2cm}|p{1.5cm}|p{2cm}p{2cm}|p{0.5cm}|p{3.0cm}|p{4.0cm}}
	author&infection&method&term for CT&CT delay&remarks&outcome(w.r.t. CT)\\
	\hline 
			Eames 2007~\cite{eames2007contact}&Theory&pair approx.\& stoch sim.&direct formulation& - &based on Eames~\cite{eames2002}&recursive CT much more effective than one-step CT; “targeted CT”: focus on core groups\\
			House et al.\ 2010~\cite{house2010}&Theory&pair approx.\& stoch sim.&direct formulation& - &based on Eames~\cite{eames2002}; focus on different social graphs (small world, scale-free)&CT has higher efficiency in clustered contact graphs\\
			Huerta et al.\ 2002~\cite{Huerta2002}&Theory&pair approx.\& stoch sim.&direct formulation& - &develop pair approximation for CT&rewiring of contact network decreases CT\\
			Tsimering et al.\ 2003~\cite{Tsimring2003}&Theory&pair approx.\& stoch sim.&direct formulation& - &based on Huerta~\cite{Huerta2002}&rewiring of contact network decreases CT\\
			\hline 
			Armbruster et al.\ 2007~\cite{Armbruster2007}&Theory&phenomen. / ODE&linear removal term& - &addresses a cost-efficiency analysis&CT only cost efficient if prevalence is small\\
			Armbruster et al.\ 2007~\cite{Armbruster2007a}&Theory&phenomen. / ODE&linear removal term& - &based on Armbruster~\cite{Armbruster2007}&CT only cost efficient if prevalence is small\\
			Mizumoto et al.\ 2013~\cite{mizumoto2013}&Theory&phenomen. / next generation operator& reduction of R0 by a factor& - &focus on the onset; multitype-branching process, analyzed by generating functions&Reff, probability for extinction, duration of a minor outbreak\\
		\end{tabular}
	\end{adjustbox}
\end{table}

\end{appendix}

\end{document}